\documentstyle[12pt,aaspp4]{article}
\def\apj #1 #2 #3 {#1, ApJ, #2, #3}                     
\def\apjl #1 #2 #3 {#1, ApJ, #2, L#3}   
\def\apjs #1 #2 #3 {#1, ApJS, #2, #3}   
\def\mnras #1 #2 #3 {#1, MNRAS, #2, #3}         
\def\pra #1 #2 #3 {#1, Phys.~Rev.~D., #2, #3}
\def\prb #1 #2 #3 {#1, Phys.~Rev.~D., #2, #3}   
\def\prc #1 #2 #3 {#1, Phys.~Rev.~D., #2, #3}
\def\prd #1 #2 #3 {#1, Phys.~Rev.~D., #2, #3}
\def\pre #1 #2 #3 {#1, Phys.~Rev.~D., #2, #3}   
\def\prl #1 #2 #3 {#1, Phys.~Rev.~Lett., #2, #3}        
\def\plb #1 #2 #3 {#1, Phys.~Lett.~B., #2, #3}
\def\science #1 #2 #3 {#1, Science., #2, #3}   
\def\nature #1 #2 #3 {#1, Nature., #2, #3}   
\def\nphysa #1 #2 #3 {#1, Nucl.~Phys.~A., #2, #3}       
\def\nphysb #1 #2 #3 {#1, Nucl.~Phys.~B., #2, #3}       
\def\nphysbs #1 #2 #3 {#1, Nucl.~Phys.~B.~Suppl., #2 #3}
\def\eprint{Los Alamos e-Print Archive., }

\def\he#1{\hbox{${}^{#1}$He}}
\def\li#1{\hbox{${}^{#1}$Li}}
\def\be#1{\hbox{${}^{#1}$Be}}
\def\yp{\hbox{$Y_{\rm p}$}}

\def\omegab{\hbox{$\Omega_b$}}

\def\mev{\mbox{~MeV}}

\def\la{\mathrel{\mathpalette\fun <}}
\def\ga{\mathrel{\mathpalette\fun >}}
\def\fun#1#2{\lower3.6pt\vbox{\baselineskip0pt\lineskip.9pt
  \ialign{$\mathsurround=0pt#1\hfil##\hfil$\crcr#2\crcr\sim\crcr}}}

\begin{document}

\bigskip
\bigskip
\medskip
\rightline{NAOJ-Th/}
\rightline{astro-ph/-----}

\title{\bf Geometrical Effects of Baryon Density Inhomogeneities on 
 Primordial Nucleosynthesis}

\author{M. Orito\altaffilmark{1,5}, T. Kajino\altaffilmark{1,2,5}, R. N. 
Boyd\altaffilmark{3}
 and G. J. Mathews\altaffilmark{4}}
\altaffiltext{1}{Department of Astronomical Science, The Graduate University 
for Advanced Studies, Mitaka, Tokyo 181, Japan}
\altaffiltext{2}{Division of Theoretical Astrophysics,
National Astronomical Observatory, Mitaka, Tokyo 181, Japan}
\altaffiltext{3}{Department of Physics, Department of Astronomy, Ohio 
State University, Columbus, OH 43210}
\altaffiltext{4}{Department of Physics, University of Notre Dame, Notre Dame, 
IN 46556}
\altaffiltext{5}{Institute for Nuclear Theory, University of Washington,
Seattle, WA 98195}

\begin{abstract}
We discuss effects of fluctuation geometry on primordial nucleosynthesis.
For the first time we consider condensed cylinder and cylindrical-shell 
fluctuation geometries in addition to condensed spheres and spherical shells.
We find that a cylindrical shell geometry allows for an appreciably 
higher baryonic contribution to be the closure density ($\Omega_b h_{50}^2
\la 0.2$) than that allowed 
in spherical inhomogeneous or standard homogeneous big bang models. 
This result, which is contrary to some other recent studies, is due to both 
geometry and recently revised estimates of the uncertainties in the
observationally inferred primordial light-element abundances. 
We also find that inhomogeneous primordial nucleosynthesis in  the 
cylindrical shell geometry can lead to significant Be and B production. 
In particular, a primordial beryllium 
abundance as high as [Be] =\,12\,+\,log(Be/H) $\approx -3$  is possible
while still satisfying all of the light-element abundance constraints.    
\end{abstract}

\keywords{ cosmology: theory - dark matter - early universe - nuclear 
reactions, nucleosynthesis, abundances}

\section{INTRODUCTION} 
The analysis of primordial nucleosynthesis provides valuable limits on 
cosmological and particle physics parameters through a comparison 
between the predicted and inferred primordial abundances of D, 
\he3, \he4, and \li7. For standard homogeneous big bang nucleosynthesis (HBBN)
the predicted primordial abundances of these 
light-elements are in accord with the value inferred from observation 
provided that baryon-to-photon ratio ($\equiv \eta$) is between about 
$2.5  \times 10^{-10}$ and $6 \times 10^{-10}$. This corresponds to an 
allowed range for the baryon fraction of the universal closure density 
$\Omega_b^{\rm HBBN}$ (\cite{walker91}; \cite{smith93}; 
\cite{copi95}; \cite{ScMa95}),
\begin{equation}
0.04 \la  \omegab^{\rm HBBN}\,h_{50}^{2} \la 0.08 ,
\label{eq:1}    
\end{equation}
where $\eta = 6.6  \times 10^{-9} \Omega_b\,h_{50}^2$. 
The lower limit on $\omegab^{\rm HBBN}$ arises mainly from the 
upper limit on the deuterium plus $^3$He abundance (\cite{yang84}; 
\cite{walker91}; \cite{smith93}), and the upper limit to $\omegab$ 
arises from the upper limit on the \he4 mass fraction $\yp$ and/or the 
deuterium abundance D/H $\ge 1.2 \times 10^{-5}$ (\cite{linsky93},\ 
\cite{linsky95}). Here, $h_{50}$ is the Hubble constant in units of 50 km s
$^{-1}$ Mpc$^{-1}$. The fact that this range for $\omegab\,h_{50}^2$ is 
so much greater than the current upper limit to the contribution from 
luminous matter $\omegab^{Lum} \la 0.01$ (see however \cite{jedamzik95}) 
is one of the strongest arguments for the existence of baryonic dark matter.

Over the years HBBN has provided strong support for the standard,
hot big bang cosmological model as mentioned above. However, as the 
astronomical data have become more precise in recent years, a possible 
conflict between the predicted abundances of the light element isotopes 
from HBBN and the abundances inferred from observations has been suggested 
(\cite{olivestei95}; \cite{steigman96a}; \cite{turner96}; \cite{hata96}; 
see also \cite{hata95}). 

There is now a good collection of abundance 
information on the \he4 mass fraction, \yp, O/H, and N/H in over 50 
extragalactic HII regions (\cite{pagel92}; \cite{pagel93}; \cite{izatov94}; 
\cite{skillman95}). In an extensive study based upon these observations, 
the upper limit to $\eta$ from the observed \he4 abundance was found to be 
$\sim 3.5 \times 10^{-10}$ (\cite{olivestei95}; \cite{olive96}) 
when a systematic 
error in $\yp$ of $\Delta Y_{sys} = 0.005$ is adopted.  Recently, it has been 
recognized that the $\Delta Y_{sys}$ may even be factor of 2 or 3 larger 
(\cite{thuan96}; \cite{copi95}; \cite{ScMa95}; \cite{sasselov95}), making
the upper limit to $\eta$ as large as $7 \times 10^{-10}$. 

On the other hand, the lower bound to $\eta$ has been derived
directly from the upper 
bound to the combined abundances of D and \he3. This is because 
it is believed that deuterium is largely converted into \he3 in stars; 
the lower bound then applies if, as has generally been assumed, 
a significant fraction of \he3 survives stellar processing (\cite{walker91}).

However, 
there is mounting evidence that low mass stars destroy \he3
(\cite{wasserburg95}; \cite{charbonnel95}),
although it is possible that massive stars produce \he3.
Therefore, the uncertainties of chemical evolution models render it  difficult 
to infer the primordial deuterium and \he3 abundances by using observations 
of the present interstellar medium (ISM) or from the solar meteoritic 
abundances. Recent data and analysis lead to a lower bound of 
$\eta\,\ga\,3.5\,\times\,10^{-10}$ on the basis of D and \he3 
(\cite{dearborn96}; \cite{hata96}; \cite{steigman96a}; \cite{steigman95}), 
if the fraction of \he3 that survives stellar processing in the course of 
galactic evolution exceeds $1/4$. This poses a potential conflict 
between the observation ($\yp$ with low $\Delta Y_{sys}$, D) and HBBN. 

In this context, possible detections (\cite{songaila94}; 
\cite{carswell94}; \cite{carswell96}; \cite{tytler94}; 
\cite{tytler96}; \cite{rugers96a},1996b; \cite{wampler96}) of an 
isotope-shifted Lyman-$\alpha$ absorption line at high redshift ($z\,
\ga\,3$) along the line of sight to quasars are of considerable 
interest. Quasar absorption systems can sample low metallicity gas at early 
epochs where little destruction of D should have occurred. Thus, they should 
give definitive measurements of the primordial cosmological D abundance. 
A very recent high resolution detection by Rugers \& Hogan (1996a) suggests 
a ratio D/H of 
\begin{equation}
          \rm{D}/\rm{H} = 1.9 \pm 0.4 \times 10^{-4}.   
        \label{eq:2}
\end{equation}
This result is consistent with the estimates made by Songaila  
et al.~(1994) and Carswell et al.~(1994), using lower resolution. 
It is also similar to that found recently in another absorption system by 
Wampler et al.~(1996), but it is inconsistent with high resolution studies
in other systems at high redshift (Tytler, Fan \& Burles 1996; Burles \&
Tytler 1996) and with the local observations of 
D and \he3 in the context of conventional models   of stellar and 
Galactic evolution (\cite{edmunds95}; \cite{gloeckler96}). 
If the high value of D/H is taken to be the primordial 
abundance, then the consistency between the observation and HBBN is 
recovered and the allowed range of $\omegab$ inferred from HBBN changes 
to $\omegab^{\rm HBBN}\,h_{50}^{2} = 0.024 \pm 0.002$ 
(\cite{jedamzik94a}; \cite{krauss94a}; \cite{vangioni95}). 
In this case, particularly if $h_{50}$ is greater than $\sim 1.5$, the
big bang prediction could be so close to the baryonic density in luminous
matter that little or no baryonic dark matter is required 
(\cite{persic92}; \cite{jedamzik95}).  
This could be in contradiction with observation, particularly if the recently
detected microlensing events (Alcock et al.~1993, 1994, 1995abc; \cite{aubourg93})
are shown to be baryonic.  This low baryonic density limit would also
be contrary to evidence (\cite{white93}; \cite{white95})
that baryons in the form of hot X-ray gas 
may contribute a significant fraction of the closure density 
 
The observations by Tytler, et al.~(1996) and Burles \& 
Tytler (1996) yield a low value of D/H. Their average abundance is 
\begin{equation}
           \rm{D}/\rm{H} = 2.4 \pm 0.9 \times 10^{-5},
        \label{eq:3}
\end{equation}
with $\pm 2\sigma$ statistical error and 
$\pm 1\sigma$ systematic error. This value is consistent with the 
expectations of local galactic chemical evolution. 
However, this value would imply 
an HBBN helium abundance of $\yp = 0.249 \pm 0.003$ which is only marginally 
consistent with the observationally inferred $\yp$ even if the high $\Delta Y_{sys}$ is 
adopted.

With this in mind, it is worthwhile to consider alternative 
cosmological models. One of the most widely investigated 
possibilities is that of an inhomogeneous density distribution 
at the time of nucleosynthesis.  Such studies were 
initially motivated by speculation (\cite{witten84}; \cite{applegate85}) 
that a first order quark-hadron phase transition (at $T \sim 100 
\mev$) could produce baryon inhomogeneities as baryon number was 
trapped within bubbles of shrinking quark-gluon plasma. In previous 
calculations using the baryon inhomogeneous big bang nucleosynthesis 
(IBBN) model, it has been usually assumed that the geometry of baryon density 
fluctuations is approximated by condensed spheres. Such geometry might be 
expected to result from a first order QCD phase transition in the limit that
the surface tension dominated the evolution of shrinking bubbles of 
quark-gluon plasma. However, the surface tension may not be large 
(\cite{kajantie90}, 1991, 1992) during the QCD transition, which could lead to a "shell" 
geometry or the development of dendritic fingers (Freese  \& Adams 1990). 
Furthermore, such fluctuations might have been produced by a number 
of other processes operating in the early universe (cf. \cite{malaney93}),
for which other geometries may be appropriate, e.g. strings, sheets, etc. 
Thus, the shapes of any cosmological baryon inhomogeneities must be 
regarded as uncertain. 
 
The purpose of this paper is, therefore, to explore the sensitivity of the 
predicted elemental abundances in IBBN models to the geometry of the 
fluctuations. We consider here various structures and profiles for the 
fluctuations in addition to condensed spheres.  Mathews et al.~
(1990, 1994, 1996) found that placing the fluctuations in spherical 
shells rather than condensed spheres allowed for lower calculated 
abundances of \he4 
and \li7 for the same $\omegab$, and that a condensed 
spherical geometry is not necessarily the optimum. Here we show that 
a cylindrical geometry also allows for an even  higher baryonic contribution 
to the closure density than that allowed by the usually adopted 
condensed sphere.  It appears to be a general result that shell geometries
allow for a slightly higher baryon density.  This we attribute
to the fact that, for optimum parameters, shell geometries involve
a larger surface area to volume ratio and hence more efficient neutron
diffusion.

An important possible consequence of baryon inhomogeneities at the time 
of nucleosynthesis may be the existence of unique nucleosynthetic 
signatures.  Among the possible observable signatures of baryon 
inhomogeneities already pointed out in previous works are the high 
abundances of heavier elements such as beryllium and boron 
(\cite{boyd89}; 
\cite{kajino90a}; \cite{malaney89a}; Terasawa \& Sato 1990; 
\cite{kawano91}), intermediate mass elements (\cite{kajino90b}), or 
heavy elements (\cite{malaney88}; \cite{applegate88}; \cite{rauscher94}).  
Such possible 
signatures are also constrained, however, by the light-element 
abundances. It was found in several previous calculations that the possible 
abundances of synthesized heavier nuclei was quite small 
(e.g., \cite{alcock90}; Terasawa \& Sato 1990; \cite{rauscher94}).  
We find, however, that substantial production of heavier elements 
may nevertheless be possible 
in IBBN models with cylindrical geometry.  

\section{BARYON DENSITY INHOMOGENEITIES}

After the initial suggestion (Witten 1985) of QCD motivated baryon inhomogeneities
it was quickly realized (\cite{applegate85}; \cite{applegate87}) that 
the abundances of primordial nucleosynthesis could be affected. 
A number of papers have addressed this point (\cite{alcock87}; 
\cite{applegate87}, 1988; \cite{fuller88}; 
\cite{kurki88}, 1990; \cite{terasawa89a}, \cite{terasawa89b},  
\cite{terasawa89c}, \cite{terasawa90}; \cite{kurki89}, \cite{kurki90a}; 
\cite{mathews90}, \cite{mathews93b}; \cite{mathews96}; 
\cite{jedamzik94a}; \cite{jedamzik95}; \cite{thomas94}; \cite{rauscher94}).  
Most recent studies in which the coupling between the baryon diffusion and 
nucleosynthesis has been properly accounted for (e.g., 
\cite{terasawa89a},  
\cite{terasawa89b}, \cite{terasawa89c}, \cite{terasawa90}; \cite{kurki89}, 
\cite{kurki90a}; \cite{mathews90}, \cite{mathews93b}; 
\cite{jedamzik94a}; \cite{thomas94}) have concluded that the upper limit on 
$\omegab\,h^2$ is virtually unchanged when compared to the upper 
limit on $\omegab\,h^2$ derived from standard HBBN.  It is also 
generally believed  (e.g. \cite{vangioni95}) that the same holds true 
if the new high D/H abundance is adopted. 

However, in the previous studies, it 
was usually assumed that a fluctuation geometry of centrally condensed 
spheres produces the maximal impact on nucleosynthesis.  
Here we emphasize that condensed spheres are not necessarily  
the optimal nor the most physically motivated 
fluctuation  geometry.

Several recent lattice QCD calculations (\cite{kajantie90}, 1991, 
1992; \cite{brower92}) indicate that the surface tension of nucleated 
hadron bubbles is relatively low.  In this case, after the hadron bubbles 
have percolated, the structure of the regions remaining in the quark phase 
may not form spherical droplets but
rather sheets or filaments. We do note that the significant effects on 
nucleosynthesis may require a relatively strong first order phase transition 
and sufficient surface tension to generate an optimum separation distance 
between baryon fluctuations (\cite{fuller88}). However, even if the surface 
tension is low, the dynamics of the coalescence of hadron droplets may 
lead to a large separation between regions of shrinking 
quark-gluon plasma.  Furthermore, even though lattice QCD has not 
provided convincing evidence for a strongly first order QCD phase 
transition (e.g., \cite{fukugita91}), the order of the transition must 
still be considered as uncertain (\cite{gottlieb91}; \cite{petersson93}). 
It depends sensitively upon the number of light quark flavors. The 
transition is first order for three or more light flavors and second 
order for two. Because the $s$ quark mass is so close to the transition 
temperature, it has been difficult to determine the order of transition. 
At least two recent calculations (\cite{iwasaki95}; \cite{kanaya96}) indicate a clear 
signature of a first order transition when realistic $u, d, s$ quark 
masses are included, but others indicate either second order or no 
phase transition at all. 

In addition to the QCD phase transition, there remain a number of 
alternative mechanisms for generating baryon inhomogeneities
 prior to the nucleosynthesis epoch (cf.~\cite{malaney93}), such as 
electroweak baryogenesis (\cite{fuller94}), inflation-generated 
isocurvature fluctuations (\cite{dolgov93}), and kaon condensation 
(\cite{nelson90}). Cosmic strings might also 
induce baryon inhomogeneities through electromagnetic (\cite{malaney89b}) 
or gravitational interactions.

Since the structures,  shapes, and origin  of any baryon 
inhomogeneities are uncertain,  
a condensed spherical geometry is not necessarily the most 
physically motivated choice.  Indeed, we will show that a condensed spherical geometry 
is also not necessarily the optimum to allow for the highest values 
for $\omegab$ while still satisfying the light-element abundance 
constraints.  Here we consider the previously unexplored cylindrical geometry.
String geometries may naturally result from various baryogenesis scenarios such as
superconducting axion strings or cosmic strings. Also, the fact that
QCD is a string theory may predispose QCD-generated fluctuations 
to string-like geometry (\cite{kajino93}; \cite{tassie93}).
Hence, cylindrical fluctuations may be a natural choice.

\section{OBSERVATIONAL CONSTRAINTS}
We adopt the following constraints on the 
observed helium mass fraction $\yp$ and \li7 taken from 
Balbes et al.~(1993), Schramm \& Mathews (1995), Copi et al.~(1995) and Olive (1996):
\begin{equation}
        0.226 \leq \yp \leq 0.247, 
        \label{eq:4}
\end{equation}
\begin{equation}
        0.7 \times 10^{-10} \leq \li7/\rm{H} \leq 3.5 \times 10^{-10}. 
        \label{eq:5}
\end{equation}

This primordial $\he4$ abundance constraint includes a 
statistical uncertainty of $\pm 0.003$ and possible systematic errors 
as much as $+0.01/-0.005$ with central value of $0.234$.
A recent reinvestigation (with new data) of the linear regression 
method for estimating the primordial $\he4$ abundance has called into 
question the systematic uncertainties assigned to $\yp$ 
(\cite{izatov96}). Our adopted upper limit to $\yp$ of Eq.~(\ref{eq:4}) is 
essentially equal to the limit derived in their study with $1\sigma$ 
statistical error.

The upper limit to the lithium abundance adopted here includes the 
systematic increase from the model atmospheres of Thorburn (1994) 
and the possibility of as much as a factor of 2 increase due to stellar 
destruction.  This is consistent with the recent observations of \li6 in
halo stars (\cite{smith92}; \cite{hobbs94}).
We note that recent discussion of model atmospheres (\cite{kurucz95}) 
suggests that as much as an order magnitude upward shift in the 
primordial lithium abundance could be warranted due to the tendency 
of one-dimensional models to underestimate the ionization of lithium. 
Furthermore,  a
recent determination of the lithium abundance 
in the globular cluster M92 having the metallicity [Fe/H] = -2.25
has indicated that at least one star out of seven shows 
[Li] =\,12\,+\,log(Li/H) $\approx 2.5$ (\cite{boesgaard96b}).
Since the abundance measurement of the globular cluster stars is 
more reliable than that of field stars, this detection along with the
possible depletion of lithium in stellar atmospheres suggests 
that a lower limit to the primordial abundance is
$ 3.2 \times 10^{-10} \leq \li7/\rm{H}$.

There also remains the question as to why several stars which are 
in all respects similar to the other stars in the Population~II
`lithium plateau', are so lithium rich or lithium deficient 
(\cite{deliyannis96}; \cite{boesgaard96a}, 1996b).
Until this is clarified, it may be premature to 
assert that the Population~II abundance of lithium reflects the primordial 
value.  The primordial abundance may instead correspond to the much higher 
value observed in Population~I stars which has been depleted down to the 
Population~II lithium plateau. The observational evidence (Deliyannis,
Pinsonneault \& Duncan 1993) for a $\pm\,25\,\%$ 
dispersion in the Population~II lithium plateau is consistent with 
this hypothesis (\cite{deliyannis93}; \cite{charbonnel95}; \cite{steigmanli7}). 
Rotational depletion was studied in detail by Pinsonneault et al.~
(1992) who note that the depletion factor could have been as large 
as $10$. Chaboyer and Demarque (1994) also demonstrated that models 
incorporating both rotation and diffusion provide a good match to 
the observed \li7 depletion with decreasing temperature in 
Population~II stars and their model indicated that the initial 
lithium abundance could have been as high as 
$\li7/\rm{H} = 1.23 \pm\,0.28 \times 10^{-9}$.

A recent study (\cite{ryan96}), which includes new data on 7 halo 
dwarfs, fails to find evidence of significant depletion through 
diffusion, although other mechanisms are not excluded. For example, 
stellar wind-driven mass loss could deplete a high primordial lithium 
abundance of down to the Population~II value [Eq.~(\ref{eq:5})] in a 
manner consistent with \li6 observations (\cite{vauclair95}). 
Furthermore, it could be possible (\cite{yoshii95}) that some of the \li6 
is the result of more recent accretion of interstellar material that 
could occur as halo stars episodically plunge through the disk. Such a 
process could mask the earlier destruction of lithium.
For comparison, therefore, we adopt a conservative upper limit 
on the primordial lithium abundance of 
\begin{equation}
        \li7/\rm{H} <   1.5 \times 10^{-9}.
        \label{eq:6}
\end{equation}

Finally,  the primordial abundance of deuterium is even harder to  
clarify since it is easily destroyed in stars (at temperatures 
exceeding about $6 \times 10^{5}$K). Previously, limits on the 
deuterium (and\ also the\ \he3) abundances have been inferred from 
their presence in 
presolar material (e.g.,~\cite{walker91}). It is also inferred from 
the detection in the local interstellar medium (ISM) through its 
ultraviolet absorption lines in stellar spectra (\cite{mccullough92}; 
\cite{linsky93},\ 1995). The limit from ISM data is consistent with 
that from abundances in presolar material. It has been argued that 
there are no important astrophysical sources of deuterium 
(\cite{epstein76}) and ongoing observational attempts to detect 
signs of deuterium synthesis in the Galaxy are  so far consistent with  
this hypothesis (see~\cite{pasachoff89}). If this is indeed so, then the 
lowest D abundance observed today should provide a lower bound to the 
primordial abundance. Recent precise measurements by Linsky et al.~
(1995, 1993) using the {\it Hubble Space Telescope} implies 
\begin{equation}
        \rm{D}/\rm{H} > 1.2 \times 10^{-5}.
        \label{eq:7}
\end{equation}
We adopt  this as a lower limit to the primordial deuterium abundance
for the purposes of exploring the maximal cosmological impact from IBBN.
In addition, we consider the two possible detections of the
deuterium abundance along the line of sight to high red shifted quasars,
Eqs.~(\ref{eq:2}) and (\ref{eq:3}) as possible limits.

In order to derive a lower limit to $\omegab\,h_{50}^2$, it is useful 
to consider the sum of deuterium plus \he3. In the context of a 
closed-box instantaneous recycling approximation, it is straightforward 
(\cite{olive90}) to show that the sum of primordial deuterium and \he3 
can be written
\begin{equation}
y_{23p} \le A_\odot^{(g_3 - 1)}y_{23\odot} \biggl({ X_\odot
\over X_p} \biggr)
             \label{eq:8}
\end{equation}
where $A_\odot$ is the fraction of the initial primordial deuterium still 
present when the solar system formed, $g_3$ is the fraction of \he3 
that survives incorporation into a single generation of stars, $y_{23\odot}$ 
is the presolar value of  [D$+$\he3]/H inferred from the 
gas rich meteorites, and $X_\odot/X_p$ is the ratio of the
presolar hydrogen mass fraction to the primordial value.
These factors together imply an upper limit (\cite{walker91}; 
\cite{copi95}) of 
\begin{equation}
y_{23p} \le 1.1 \times 10^{-4}.
 \label{eq:9}
\end{equation}

\section{CALCULATIONS}
The calculations described here are based upon the coupled diffusion and
nucleosynthesis code of Mathews et al.~(1990), but with a number of
nuclear reaction rates updated and the numerical diffusion scheme
modified to accommodate cylindrical  geometry.  We also have implemented an
improved numerical scheme which gives a more accurate description of
the effects of proton and ion diffusion,  and Compton drag
at late times.  Although our approach is not as sophisticated as that 
of Jedamzik et al.~(1994a), it produces essentially the same
results for the  parameters employed here. We have also included all of the
new nuclear reaction rates summarized in  Smith et al.~(1993) as 
well as those given in Thomas et al.~(1993).
We obtain the same result as Smith et al.~(1993) 
using these rates and homogeneous conditions in  our IBBN model 

Calculations were performed in a cylindrical geometry both with the high 
density regions in the center (condensed cylinders), and with the high 
density regions in the outer zone of computation (cylindrical shells).
Similarly, calculations were made in a spherical geometry with the high density 
regions in the center (condensed spheres) and with the high density region 
in the outer zones of computation (spherical shells). 

In the calculations, the fluctuations are resolved into 16 zones of 
variable width as described by Mathews et al.~(1990). We assumed 
three neutrino flavors and an initially
 homogeneous density within the fluctuations. Such 
fluctuation shapes are the most likely to emerge, for example, after 
neutrino-induced expansion (\cite{jedamzik94}). We use a neutron 
mean life-time of $\tau_{n} = 887.0$ (\cite{particle94}). In addition to 
the cosmological parameter, $\omegab$ and fluctuation geometry, there 
remain three parameters to specify the baryon inhomogeneity. They are: 
$R$, the density contrast between the high and low-density regions; 
$f_{v}$, the volume fraction of the high-density region; and $r$, the 
average separation distance between fluctuations. 

\section{RESULTS}
The parameters $R$ and $f_{v}$ were optimized to allow for the highest 
values for $\omegab\,h_{50}^{2}$ while still satisfying the 
light-element abundance constraints. For fluctuations represented by 
condensed spheres, optimum parameters are $R \sim 10^{6}$ and 
$f_v^{1/3}  \sim 0.5$ (\cite{mathews96}).  For other fluctuation 
geometries, we have found that optimum parameters are:
\begin{mathletters}
\begin{eqnarray*}
R &\sim & 10^{6};  \qquad \qquad \mbox{for all fluctuation geometries} 
\\
\\
f_{v}^{1/3} &\sim & 0.19;  \qquad \quad \mbox{~~} \mbox{for spherical shells} 
\\
\\
f_{v}^{1/2} &\sim &\left\{
\begin{array}{rl}
0.5;& \quad \quad \mbox{for condensed cylinders} \\
0.15;& \quad \quad \mbox{for cylindrical shells}, 
\end{array}\right.
\end{eqnarray*}
\mbox{although there is not much sensitivity to $R$ once 
$R\,\ga\, 10^3$.}
\end{mathletters}
Regarding $f_v$, we have written 
the appropriate length scale of high density regions, i.e.
$f_v^{1/3}$ and $f_v^{1/2}$ for the spherical and cylindrical
fluctuation geometries, respectively.
The variable parameters in the calculation are then the fluctuation 
cell radius $r$, and the total baryon-to-photon ratio $\eta$ (or 
$\omegab\,h_{50}^{2}$).

\subsection{Constraints on \omegab$h_{50}^2$}
Figures~\ref{fig:1}, \ref{fig:2}, \ref{fig:3a}, and \ref{fig:4a} show contours 
of allowed parameters 
in the $r$ versus $\eta$ and $r$ versus $\omegab\,h_{50}^{2}$ plane for 
the adopted light-element abundance constraints of Eqs.~(\ref{eq:4}) - 
(\ref{eq:6}) and for a possible Lyman-$\alpha$ D/H of Eqs.~(\ref{eq:2}) 
and (\ref{eq:3}), for the condensed sphere, spherical shell, 
condensed cylinder, and cylindrical shell fluctuation geometries, 
respectively. 
The fluctuation cell radius $r$ is given in units of meters for a comoving 
length scale fixed at a temperature of $kT = 1 \mev$. 
Both of the possible 
\li7 limits, Eqs.~(\ref{eq:5}) and (\ref{eq:6}) which we have discussed above,
are also drawn as indicated.  
In order to clearly distinguish the two abundance constraints,
we use the single and double-cross hatches for the regions allowed 
by the adopted lower (Eq.~(\ref{eq:5})) and higher (Eq.~(\ref{eq:6}))
limits to the \li7 primordial abundance.

Even in the IBBN scenario, if the low D/H of Eq.~(\ref{eq:3}) 
(\cite{burles96}) is adopted as primordial, this range for D/H appears 
to be compatible with the \li7 abundance only when a higher (Population~I) 
primordial \li7 abundance limit is adopted, 
except for a very narrow region of $\eta \sim 6 \times 10^{-10}$ and
$r \leq 10^{2}\,m$.
This conclusion remains unchanged for any other fluctuation
geometries. Therefore, the acceptance of the low 
(Burles \& Tytler 1996) value of D/H would 
strongly suggest that significant depletion of \li7 has occurred.

In contrast, adoption of the high D/H of Eq.~(\ref{eq:2}) 
(\cite{rugers96a}) as primordial allows the concordance of all 
light-elements. The upper limits to $\eta$ and $\omegab\,h_{50}^{2}$  
are largely determined by D and $\li7$. The concordance range for the 
baryon density is comparable to that for HBBN for small separation distance 
$r$.  However, there exist other regions of the parameter space with optimum 
separation distance, which roughly corresponds to the neutron diffusion length 
during nucleosynthesis (\cite{mathews90}), with an increased maximum allowable 
value of the baryonic contribution to the closure density to 
$\omegab\,h_{50}^{2} \leq 0.05$ for the cylindrical geometry,
as displayed in Fig.~\ref{fig:4a}.
This is similar to the value for spherical shells as shown in 
Mathews et al.~(1996) and also in Fig.~\ref{fig:2} in the present work.
The condensed sphere limits, however, are 
essentially unchanged from those of the HBBN model.
If the primordial \li7 abundance could be as high as the upper limit 
of $\mbox{Li/H} \leq 1.5 \times 10^{-9}$, the maximum allowable 
value of the baryonic content for the condensed sphere would increase to 
$\omegab\,h_{50}^{2} \leq 0.08$, with similar values 
for the spherical shell (\cite{mathews96}). For both the condensed 
cylinders and cylindrical shells, the upper limits could be as high as 
$\omegab\,h_{50}^{2} \leq 0.1$ as shown in Figs.~\ref{fig:3a} and 
\ref{fig:4a}. These higher upper limits relative to those of the HBBN are of interest 
since they are consistent with the inferred baryonic mass in the form of 
hot X-ray gas (\cite{white93}; \cite{white95}) in dense galactic clusters. 
The acceptance of this consistence, as noted above, requires the significant 
stellar depletion of \li7.

In Figures~\ref{fig:3b} and \ref{fig:4b}, we also show contours for the 
condensed cylinder and cylindrical shell geometries, respectively, but this time
with the conventional light-element constraints of 
Eqs.~(\ref{eq:4}), (5), (7), and (9) 
as indicated. Since the results  for the condensed sphere and 
spherical shell geometries with this set of the conventional abundance 
constraints have already been discussed by 
Mathews et al.~(1996), we do not show those contours here.
The cylindrical shell geometry of the present work gives the highest 
allowed value of $\omegab\,h_{50}^{2}$.
Figure~\ref{fig:4b} shows that the 
upper limits to $\eta$ and $\omegab\,h_{50}^{2}$  are largely determined 
by $\yp$ and $\li7$. 
The upper limits for a cylindrical shell geometry could be as high as 
$\omegab\,h_{50}^{2} \leq 0.13$ with similar results for the spherical 
shell geometry (\cite{mathews96}). A high primordial lithium abundance 
would increase the allowable baryonic content to  as high as 
$\omegab\,h_{50}^{2} \leq 0.2$.  The reason that shell geometries allow for 
higher baryon densities we attribute to more efficient neutron diffusion which
occurs when the surface area to volume area is increased.  This
allows for more initial diffusion to produce deuterium, and
more efficient back diffusion to avoid over producing \li7.

\subsection{Observational Signature}
The production of beryllium and boron 
as well as lithium in IBBN models can be sensitive 
to neutron diffusion. Therefore, their predicted abundances 
are sensitive to not only the fluctuation parameter $r$, $R$, and 
$f_{v}$ but also the fluctuation geometry (\cite{boyd89}; 
\cite{malaney89a}; \cite{kajino90a}; Terasawa \& Sato 1990). 
Figures~\ref{fig:5} - \ref{fig:7} show the contours of the 
calculated abundances for lithium, beryllium and boron, respectively 
in the $r$ versus $\eta$ (and $r$ versus $\omegab~h_{50}^{2}$) plane.
the shaded region depict is allowed values of $r$ and $\eta$
from the light element 
abundance constraints [cf. Fig.~\ref{fig:4b}] for a cylindrical
shell fluctuation geometry.
The contour patterns of lithium (Fig.~\ref{fig:5}) and boron 
(Fig.~\ref{fig:7})
abundances are very similar, whereas there is no similarity found
between lithium (Fig.~\ref{fig:5}) and beryllium (Fig.~\ref{fig:6}) abundances. 

In order to understand the similarities and differences among
these three elemental abundances, we show in  
Figs.~\ref{fig:8} and \ref{fig:9} the decompositions of 
the A = 7 abundance into \li7 and \be7 and the boron abundance
into $^{10}$B and $^{11}$B.
These Figures show also the 
dependence of the predicted LiBeB abundances in IBBN on the scale of 
fluctuations for a cylindrical shell geometry with fixed 
$\omegab\,h_{50}^{2} = 0.1$.  This value of 
$\omegab\,h_{50}^{2}$ corresponds to a typical value in the allowable 
range of $\eta$ in Fig.~\ref{fig:4b}, which optimizes
the light element abundance constraints, even 
satisfying the lower \li7 abundance limit of Eq.~(\ref{eq:5}).  
The fluctuation parameters $f_{v}$ and $R$ are
the same as in Fig.~\ref{fig:4b}.
Once the baryonic content $\omegab$ is fixed, the only variable parameter 
is the separation distance, $r$. 

As can be seen in Fig.~\ref{fig:8}, as the separation $r$ increases,      
neutron diffusion plays an increasingly important role in the production 
of $t$ and, by the \he4($t, \gamma$)\li7 reaction.
It works maximally around  
$r \sim 10^{4}$~m, which is the typical length scale of neutron 
diffusion at $kT = 1 \mev$.
A similar behavior is observed 
in the \li7($t, n$)\be9 reaction.  This reaction  produces 
 most of the \be9
in neutron rich environments where $t$ and \li7 are abundant, as 
was first pointed out by Boyd and Kajino (1989).
At other separation distances $r$ in a $\omegab\,h_{50}^{2} = 0.1$
model, most of the A = 7 nuclides are created 
as \be7 by the \he4(\he3, $\gamma$)\be7 reaction.
In the limit of $r$ = horizon scale, the nucleosynthesis
products are approximately equal to the sum of those produced
in the proton-rich and neutron-rich zones separately 
(\cite{jedamzik94b}).  
The predominant contribution from the proton-rich
zones makes the \be7 abundance almost constant at larger $r$, 
while both \li7 and \be9 decrease as $r$ increases toward
the horizon at any separation distance.

Figure~\ref{fig:9} shows that $^{11}$B is a predominant component of the total
boron abundance at any separation distance.  This is true for
almost all values $\omegab\,h_{50}^{2}$.
It has been pointed out (\cite{malaney88}; \cite{applegate88}; 
\cite{kajino90a}) that most
$^{11}$B is produced by the \li7($n, \gamma$)\li8($\alpha, n$)$^{11}$B 
reaction sequence in neutron-rich environments at relatively 
early times when most of the other
heavier nuclides are made.  Recent measurements of the 
previously unmeasured
\li7($\alpha$,n)$^{11}$B reaction cross section (\cite{boyd92}; 
\cite{gu95}; \cite{boyd96}) at the energies of cosmological interest have 
removed the significant ambiguity 
in the calculated $^{11}$B abundance due to this reaction.
The factor of two discrepancy among several different measurements of the
reaction cross section for \li7($n, \gamma$)\li8 was also
resolved by the new measurement (\cite{nagai91}). 
The \li7($\alpha, \gamma$)$^{11}$B reaction
also makes an appreciable but weaker contribution to the production of
$^{11}$B in the neutron-rich environment. 
In the proton-rich environment, on the other hand, the
\be7($\alpha, \gamma$)$^{11}$C reaction contributes largely
to the production of $^{11}$C which beta decays to $^{11}$B in 20.39 min. 
These facts explain why the contour patterns of the lithium
and boron abundances in Figs.~\ref{fig:5} and \ref{fig:7} look very similar.
 
It is conventional in the literature to quote the beryllium  and boron 
abundance relative to H $=10^{2}$. Hence, one defines the quantity 
[X] $= 12+\mbox{log(X/H)}$. 
In cylindrical shell fluctuation geometry
the beryllium abundance can take the 
value of $\mbox{[Be]} \sim -3$ while still satisfying all of the 
light-element abundance constraints and the Population~II lithium 
abundance constraint 
(Figs.~\ref{fig:5} and \ref{fig:6}).  This abundance is higher by three orders 
magnitude than 
that produced in the HBBN model with conventional light-element 
abundance constraints. This result is contrary to a recent result 
with the condensed sphere geometry and for a more 
restricted parameter space (\cite{thomas94}).
Recent beryllium observation of Population~II stars (\cite{rebolo88}; 
\cite{ryan90},\ 1992; \cite{ryan96a}; \cite{gilmore92a},\ 1992b; 
\cite{boesgaard93}; \cite{boesgaard94},\ 1996a,b) have placed the upper 
limit on the primordial \be9 abundance to $\mbox{[Be]} \sim -2$,
one order magnitude greater than the beryllium abundance 
in the IBBN cylindrical model.

The calculated boron abundance at the optimum separation distance is 
essentially equal to the value of the HBBN model.  
However, a high primordial lithium 
abundance would increase the upper limit to $\omegab\,h_{50}^{2}$. 
In this case, the boron 
abundance could be one or two orders magnitude larger than that 
of the HBBN model (Fig.~\ref{fig:7}).

\section{CONCLUSIONS}
We have reinvestigated the upper limit to $\eta$ and 
$\omegab\,h_{50}^{2}$ in inhomogeneous primordial nucleosynthesis 
models. We have considered effects of various geometries. In particular, 
for the first time we consider cylindrical geometry. We have also 
incorporated recently revised light-element abundance constraints  including
implications of the possible detection 
(\cite{songaila94}; \cite{carswell94}; \cite{carswell96}; \cite{tytler94}; 
\cite{tytler96}; \cite{rugers96a}, 1996b; \cite{wampler96}) of 
a high deuterium 
abundance in Lyman-$\alpha$ absorption systems. We have shown that 
with low primordial deuterium (\cite{tytler94}; \cite{tytler96}),
 significant depletion of \li7 is required to obtain 
concordance between predicted light-element 
abundance of any model of BBN and the observationally inferred primordial 
abundance. If high primordial deuterium (\cite{rugers96a}) is adopted 
(Eq.~(\ref{eq:2})),  
there is a concordance range which is largely determined by D/H, and the 
upper limit to $\omegab\,h_{50}^{2}$ is 0.05.
However, with the presently adopted (Eqs.~(\ref{eq:4}), (\ref{eq:6}), 
(\ref{eq:7}), (\ref{eq:9}))
light-element abundance constraints (\cite{ScMa95}; \cite{copi95}; 
\cite{olive96}), values of $\omegab\,h_{50}^{2}$ as large as 0.2  are 
possible in IBBN models with cylindrical-shell fluctuation geometry.

We have also found that significant beryllium and boron production is possible 
in IBBN models without violating the light element abundance constraints. 
The search for the primordial abundance of these elements in low metallicity
stars could, therefore, be a definitive 
indicator of the presence or absence of cylindrical baryon inhomogeneities in the 
early universe.

\begin{figure}
\caption{Contours of allowed values for baryon-to-photon ratio $\eta$ (or
$\Omega_b h_{50}^2$) and fluctuation separation radius $r$ based upon
the various light-element abundance
constraints as indicated.  The separation $r$ is
given in units of meters comoving at $kT = 1\mev$.
This calculation is based upon
baryon density fluctuations represented by condensed spheres.
The cross hatched region is allowed by the adopted
primordial abundance limits with high (Eq.~(\protect\ref{eq:2})) and low 
(Eq.~(\protect\ref{eq:3})) deuterium abundance in Lyman limit systems
and also a higher extreme \li7 upper limit (Eq.~(\protect\ref{eq:6})).
The single hatched region depicts the allowed parameters              
for lower \li7 (Eq.~(\protect\ref{eq:5})) constraint.
Note that the \li7 abundance is the sum of \li7 and \be7.}
        \protect\label{fig:1}
\end{figure}
 
\begin{figure}
\caption{Same as Fig.~1, but for fluctuations represented
by spherical shells.}
        \protect\label{fig:2}
\end{figure}
{
 \setcounter{enumi}{\value{figure}}
 \addtocounter{enumi}{1}
 \setcounter{figure}{0}
 \renewcommand{\thefigure}{\theenumi(\alph{figure})}
\begin{figure}
\caption{Same as Fig.~1, but for fluctuations represented
by condensed cylinders.  
Adopted primordial deuterium abundance constraints 
are inferred from observations of Lyman limit systems 
(Eqs.~(\protect\ref{eq:2}) and (3)).}
      \protect\label{fig:3a}
\end{figure}
\begin{figure}
\caption{Same as Fig.~1, but for fluctuations represented by condensed 
cylinders.  Adopted primordial deuterium and $^3$He abundance constraints 
are inferred from observations of ISM (Eqs.~(\protect\ref{eq:7}) and (9)).}
      \protect\label{fig:3b}
\end{figure}
 \setcounter{figure}{\value{enumi}}
 \setcounter{enumi}{\value{figure}}
 \addtocounter{enumi}{1}
 \setcounter{figure}{0}
\begin{figure}
\caption{Same as Fig.~3(a), but for fluctuations represented
by cylindrical shells.}
        \protect\label{fig:4a}
\end{figure}
\begin{figure}
\caption{Same as Fig.~3(b), but for fluctuations represented
by cylindrical shells.}
        \protect\label{fig:4b}
\end{figure}
\setcounter{figure}{\value{enumi}}
 }
\begin{figure}
\caption{Contours of the predicted abundance of lithium (the sum of
\li7 and \be7), for baryon-to-photon ratio $\eta$  (or
$\Omega_b h_{50}^2$) and fluctuation separation radius $r$,
in the cylindrical shell fluctuation geometry.
The shaded region displays the allowed  $\eta-r$ region from Fig.~4(b).}
       \protect\label{fig:5}
\end{figure}
\begin{figure}
\caption{Same as Fig. 5, but for beryllium, \be9.}
        \protect\label{fig:6}
\end{figure}
\begin{figure}
\caption{Same as Fig.~5, but for boron, $^{10}$B + $^{11}$B.}
        \protect\label{fig:7}
\end{figure}
\begin{figure}
\caption{Lithium and beryllium abundances as function of proper 
separation distance in units of meters comoving at $kT = 1\mev$
for fixed $\Omega_b h_{50}^2 = 0.1$.
Refer to the abundance scales in $l.h.s.$ for lithium and $r.h.s.$ 
for beryllium.}
        \protect\label{fig:8}
\end{figure}
\begin{figure}
\caption{Same as Fig.~8, but for boron.}
        \protect\label{fig:9}
\end{figure}

\end{document}